\documentclass[%
 reprint,
bibnotes,
 amsmath,amssymb,
 aps,
prb,
]{revtex4-1}

\usepackage{graphicx}
\usepackage{dcolumn}
\usepackage{bm}
\usepackage{multirow}
\usepackage{tabularx}
\usepackage{array}
\usepackage[usenames,dvipsnames]{color}
\usepackage[normalem]{ulem}

\newcolumntype{M}[1]{>{\centering\arraybackslash}m{#1}}

\definecolor{green}{rgb}{0.13, 0.55, 0.13}

\begin{document}

\preprint{APS/123-QED}


\title{First-principles study of the electronic and magnetic properties of cubic GdCu compound}

\author{Vikas Kashid}
\email{v.kashid@fz-juelich.de}
\affiliation{Peter Gr\"{u}nberg Institut and Institute for Advanced Simulation, Forschungszentrum J\"{u}lich and JARA, 52425 J\"{u}lich, Germany}
\author{Ersoy \c{S}a\c{s}{\i}o\u{g}lu}
\affiliation{Institute of Physics, Martin Luther University Halle-Wittenberg, 06120 Halle (Saale), Germany}
\author{Gustav Bihlmayer}
\affiliation{Peter Gr\"{u}nberg Institut and Institute for Advanced Simulation, Forschungszentrum J\"{u}lich and JARA, 52425 J\"{u}lich, Germany}
\author{Alexander B. Shick}
\affiliation{Institute of Physics, Czech Academy of Sciences, Na Slovance 2, CZ-182 21 Prague, Czech Republic}
\author{Stefan Bl\"{u}gel}%
\affiliation{Peter Gr\"{u}nberg Institut and Institute for Advanced Simulation, Forschungszentrum J\"{u}lich and JARA, 52425 J\"{u}lich, Germany}

\date{\today}

\begin{abstract}
The structural, electronic, and magnetic properties of bulk GdCu (CsCl-type) are investigated using spin density 
functional theory, where highly localized $4f$ orbitals are treated within LDA+$U$ and GGA+$U$ methods.
The calculated magnetic ground state of GdCu using collinear as well as spin spiral calculations exhibits a C-type 
antiferromagnetic configuration representing a spin spiral propagation vector $\mathbf{Q}=\frac{2\pi}{a}(\frac{1}{2},\frac{1}{2},0)$.  
The parameters of the effective Heisenberg Hamiltonian are evaluated from a self-consistent electronic structure and are 
used to determine the magnetic transition temperature. The estimated N\'{e}el temperature of the cubic GdCu using GGA+$U$ 
and LDA+$U$ density functionals within the mean field  and random phase approximations are in good agreement 
with the experimentally measured values.  In particular, the theoretical understanding of the experimentally observed core 
Gd $4f$ levels shifting in photoemission spectroscopy experiments is investigated in detail. By employing the self-consistent
constrained random-phase approximation we determined the strength of the effective Coulomb interaction (Hubbard $U$) between 
localized $4f$ electrons. We find that, the shift of Gd-$4f$ states in GdCu with respect to bulk Gd within DFT+$U$ is sensitive 
to choice of lattice parameter. The calculations for $4f$-level shifts using DFT+$U$ methods as well as Hubbard-1 approximation 
are not consistent with the experimental findings.

\end{abstract}

\pacs{Valid PACS appear here}
\maketitle


\section{\label{sec:intr}Introduction}

The rare-earth metal Gd is known for its room temperature ferromagnetic property. Due to the localized character of $4f$ electrons in rare earth elements, which is responsible for magnetism, Gd and its various compounds are interesting candidates for unique and fascinating magnetic properties. Gd compounds are ideal sources to test various models of magnetism. During last 60 years, several attempts have been made to understand the interaction in Gd compounds responsible for magnetic ordering \cite{kirch,franse}. The open shell of strictly localized $4f$ electrons does not interact with the atoms at neighboring sites. The magnetism is coupled to the other sites through $6s$, $6p$ and $5d$ kind of conduction electrons, by so called oscillatory Ruderman-Kittel-Kasuya-Yosida (RKKY) type of exchange interaction. In Gd compounds/alloys, the exchange mechanism is mediated through the conduction electrons of different kind of atoms, which leads to an interesting type of magnetic ordering (ferro, antiferromagnetic or even more complex type due to the $4f$-state). This oscillatory exchange coupling mechanism leads to interesting magnetic structures and is of great interest for applications in intermetallics and magnetic storages. \\

The binary compounds of Gd, however, are found to exist in multiple structural as well as magnetic phases and exhibit lattice instabilities. It has been observed that the rare earth compounds formed with heavy elements crystallize in the cubic CsCl-type structure with an antiferomagnetic configuration, whereas compounds with lighter elements are ferromagnetic \cite{postnikov, blanco}. In particular, GdCu is observed to adopt the cubic structure at room temperature, however, undergoes a partial phase transformation to the orthorhombic FeB-type structure at low temperature. A neutron diffraction study by Blanco \textit{et al.} \cite{blanco} shows that both CsCl- and FeB-type of crystal structures exist in bulk GdCu samples for the range from $5$~K to $300$~K and the percentage of each phase is a function of temperature. These partial phase transitions are found to be diffusionless and displacive \cite{blanco, sathe,krystian}. However, the powdered GdCu samples have only CsCl-type of the crystal structure in the above mentioned temperature range.  The M\"{o}ssbauer absorption spectrum of powdered GdCu samples confirmed only CsCl-type of phase for temperature range from  $4.2$~K to $78$~K \cite{ross}. \\

The experimental approach to understand the electronic structure of GdCu compounds (\textit{viz.}, GdCu, GdCu$_{2}$, GdCu$_{9}$, \textit{etc.}) was carried out by Szade \textit{et al.},\cite{szade1, szade2} and Lachnitt \textit{et al.} \cite{lachnitt} using photoelectron spectroscopy. 
The authors observed that the intermetallic alloying of Gd with Cu leads to a chemical interaction in terms of charge transfer and it affects the band structure. By using the valence band spectrum of bulk Gd as a reference, the photoemission spectra showed a shifting of $4f$ levels towards higher binding energy by $0.3$~eV. Besides the Gd $4f$ level, Gd $4d$, Cu $3d$ and Cu $2p_{3/2}$ levels were observed to shift in GdCu spectra with respect to their pure elemental spectral counterparts. The shifting of the levels was explained using a charge transfer model. In this paper, we use first-principles calculations based on density-functional theory (DFT) to verify experimentally observed core level shifts of Gd and Cu atoms. When studying rare-earth magnetism, fully first-principles calculations remain a challenge due to the description of the strongly correlated nature of $4f$ electrons. Various models were presented to describe magnetism in rare-earth elements and their compounds, \textit{viz.}, $4f$-band \cite{singh2}, $4f$-core \cite{dinmmock}, hybrid and LDA+$U$ \cite{shick,shickab}. Among these methods, the ground state electronic structure and the magnetic moment were explained correctly in Gd by treating $4f$ electrons within LDA+$U$ method. In particular for GdCu, the attempts were made to investigate the magnetic ground state using the LMTO approach \cite{postnikov} and the TB-LMTO approach \cite{lachnitt}, however, the calculations failed to describe the energy position of the $4f$ states with sufficient accuracy, because the strong correlation effects in the $4f$ states were not included.  The correct energy of $4f$ states with respect to the experiment were calculated by Knyazev \textit{et al.} using TB-LMTO-ASA with LSDA+$U$ approach\cite{Knyazev}, and the study shows that the interband absorption spectra of GdCu compounds are due to electron transitions between both the spin carriers of $d$ and $p$ of Cu and between minority carriers $d$ and $f$ of Gd. However, the detailed insight into the magnetic and electronic structure is still missing. 
 
In this paper, we use the full-potential linearized augmented planewave (FLAPW) method to investigate the detailed electronic band structure of cubic GdCu. Our calculations show that the magnetic ground state is a checkerboard antiferromagnetic spin configuration in the (100) plane (type-C). Through our calculations, we shed light onto the relative shift of core states with reference to their bulk elemental counter-parts.  The paper is organized as follows: The computational methods are described in Sec.\,\ref{comp}. The general trends in the electronic and magnetic structure are discussed in Sec.\,\ref{dos-band}, the magnetic exchange parameters evaluated using collinear and non collinear magnetic states are discussed in Sec.\,\ref{sec:MIP}. The estimation of the critical temperature using the mean field approximation (MFA) and random phase approximation (RPA)  is discussed in Sec.\,\ref{sps-neel}.  
The strength of the effective Coulomb interaction (Hubbard $U$) between $4f$ electrons within the self-consistent constrained random-phase approximation (cRPA) and details of the core energy levels  based on DFT+$U$ and Hubbard-I methods are described in Sec.\,\ref{core}, followed by general conclusions in Sec.\,\ref{con}.

\section{Computational Details\label{comp}}
The present collinear and non-collinear spin-polarized calculations are performed using the FLAPW method\cite{wimmer,weinert}, as implemented in FLEUR code \cite{fleur,Kurz:04} . The planewave cut-off for these 
basis functions was set to $K_{\mathrm{max}}=4.0~\mathrm{au}^{-1}$. The charge density and potentials were 
expanded up to a cut-off $G_{\mathrm{max}}=10.7~ \mathrm{au}^{-1}$. The muffin-tin radii for Gd and Cu are 
set to $2.80~\mathrm{au}$ and $2.33~\mathrm{a.u.}$, respectively. Inside the muffin-tin sphere, the wavefunctions, 
densities and the potentials were allowed to be expanded in  spherical harmonic functions up to $l_{\mathrm{max}}=10$. 
As described for bulk Gd by Kurz \textit{et al.}\cite{kurz}, the $5s$ and $5p$ 
semi-core states are treated as valence states.  In all the calculations, the semi-core states are treated 
using local orbitals~\cite{singh,kurz}. The calculations are performed using generalized gradient approximation 
(GGA) as proposed by Perdew-Burke-Ernzerhof~\cite{perdew}, and the local density approximation (LDA) by 
Vosko-Wilk-Nusair \cite{vosko} together with their LDA+$U$ and GGA+$U$ variants. 
The LDA+$U$ and GGA+$U$ formalisms are implemented according to Shick \textit{et al.}~\cite{shick}. 

We used the $U$ parameters for bulk Gd as in Refs.\,\onlinecite{shick,kurz}, \textit{i.e.}, $U=6.7$ eV and $J=0.7$ eV for valence band 
properties, but we have finely optimized the value of $U$ in order to understand the energy of core levels of 
GdCu system. The Brillouin-zone integration is performed with $13 \times 13 \times 11$ mesh for bulk hcp Gd and a 
$7 \times 7 \times 11$ mesh of $\mathbf{k}$-points for GdCu (within C-type tetragonal unit cell) during self-consistent electronic structure relaxations. The structural 
optimization was performed using LDA+$U$ and GGA+$U$ formalism. For bulk Gd, the equilibrium lattice constant is 
achieved by keeping $c/a$ ratio fixed to the experimental value of 1.597~\cite{banister}.

\begin{figure}[b]
\includegraphics[width=0.9in,angle=90]{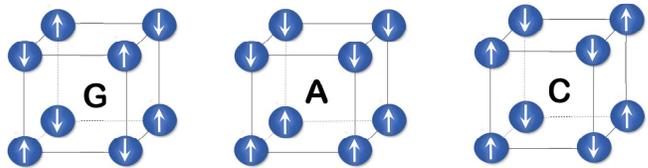}
\caption{Schematic representation of G-, A- and C-type antiferromagnetic structures. The spin orientations are shown by arrows at lattice sites. The Cu atoms are not shown for clarity.}
\label{3-antif}
\end{figure}

\section{Results and Discussion \label{result}}
\begin{figure*}[t]
\includegraphics[scale=0.4, angle=270]{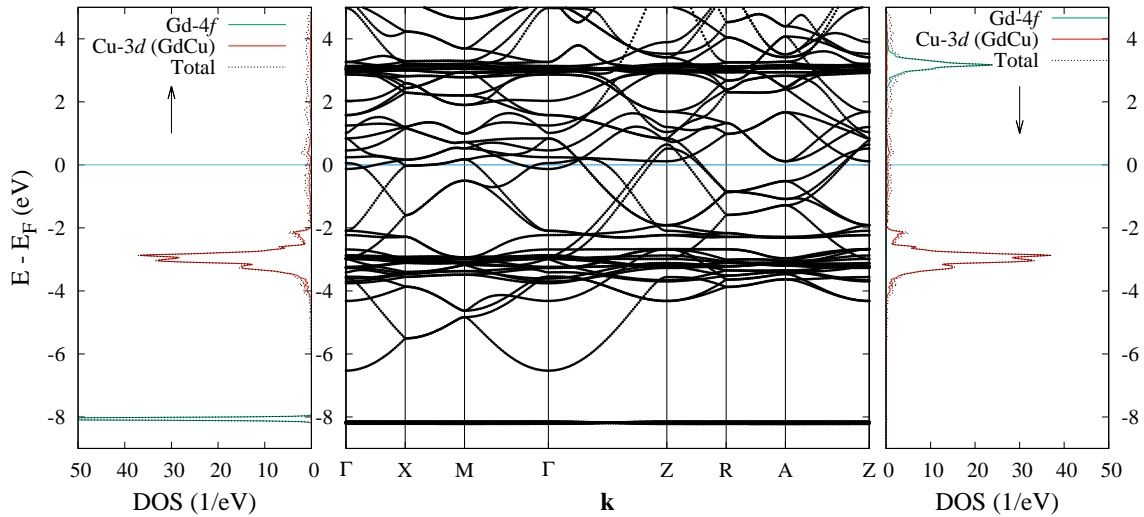}\\
\caption{GGA-PBE density of states and band structure of GdCu (C-type magnetic structure) calculated within 4$f$-core model with inclusion of Hubbard $U$. The DOS for majority spins is shown in the left side panel, whereas, the right panel depicts the minority spin DOS. The total DOS, Gd-$4f$, Cu-$3d$ states are shown for each spin. }
\label{dos-band-exlc}
\end{figure*}

This section is divided into four parts.  In the first part we discuss the general trends in the electronic band structure 
and ground state magnetic properties of GdCu. The second part deals with a detailed discussion of the exchange interactions 
and spin spirals. The third part is devoted to the discussion of the N\'eel temperature and in the last part we discuss the cRPA calcularions for the strength 
of the effective Coulomb interaction (Hubbard $U$) for $4f$ electrons and  describe the details of the calculations of the core energy levels.

\subsection{Band properties and the magnetic structure\label{dos-band}}

First we present our results for hcp Gd and cubic GdCu from our first-principles calculation and 
compare those with the previous experimental as well as theoretical findings for benchmarking.  Bulk Gd resides 
in the hexagonal closed packed (hcp) lattice (space group: $P6_{3}/mmc$, No.\ 194) with with a $c/a$ 
ratio of $1.597$, whereas GdCu has a CsCl-type cubic structure (space group: $Pm3m$, No.\ 221).

We compute the electronic and magnetic properties including the effect of Hubbard $U$ 
in the electron density functional, as discussed in Ref. \cite{kurz}. 
We use a smaller values of the $U$ from the literature  as used by Kurz \textit{et al.}, \cite{kurz} and obtained by Shick \textit{et al.}, \cite{shick}.
Furthermore, we kept the  $U$ and $J$ values unchanged for Gd and GdCu. The estimation of $U$ within cRPA method for 
$4f$-electron materials will be discussed in a later section. 

\subsubsection{Magnetic Ground State}
First we discuss the results obtained using the experimental lattice constant of $3.629$~\AA ~ and $3.502$~\AA~ for bulk Gd and GdCu, respectively. Within the LDA+$U$ and GGA+$U$ approach, the calculated magnetic ground state of bulk hcp Gd is ferromagnetic, in agreement with previous findings. The exchange splitting of $4f$ states in bulk Gd is $11.3$~eV, in good agreement with the previous theoretical result \cite{kurz} and the experimental value of 11~eV \cite{shick}. The spin-polarized calculations for cubic GdCu were performed for ferromagnetic (FM) and three types of antiferromagnetic (AFM) configurations~\cite{huai}, \textit{viz.}, G-, A- and C- types, as shown in Fig.~\ref{3-antif}. Our first-principles results demonstrate that for GdCu, the C-type antiferromagnetic configuration is energetically more favorable than the G- and A-type  states as well as  the FM state. The obtained ground state is in agreement with the experimental finding by Blanco \textit{et al.},\cite{blanco, sathe} and previous theoretical results by Knyazev \textit{et al.}~\cite{Knyazev}. The calculated energy difference between the magnetic A-, C-,  and G-type states and the FM state are collected as $\Delta E_A$, $\Delta E_C$, $\Delta E_G$ in Tab.~\ref{eqlc-fm-afm}. We notice that the choice of the exchange correlation functional, LDA$+U$ versus  GGA$+U$, has a significant influence on the relative energetics of the magnetic states.

\subsubsection{Electronic Structure}
The density of states for Gd $4f$ states and Cu $3d$ states along with the GdCu band structure for the C-type magnetic structure is shown in Fig.~\ref{dos-band-exlc}. One important feature of the band structure is the location of Gd majority $4f$ level around $8.06$~eV below the Fermi energy, which is consistent with the experimental result of $8.1$~eV by Szade \textit{et al.}, \cite{szade1}.  Similar to bulk Gd, the intra-atomic exchange interaction splits the Gd $4f$ states in GdCu by $11.23$~eV, with completely filled-up $4f$ majority states. The $4f$ bands exhibits a narrow, dispersionless behavior. As observed in the photoemission spectra of GdCu \cite{szade1}, we observe Cu $3d$ states located  approximately $3$~eV below the Fermi energy. The Cu $3d$ states in GdCu appear to be more localized than in fcc Cu. The spin up and down components are identical indicating the absence of spin splitting, as expected for an AFM structure. The more dispersive behavior in the valence band region of the band structure is due to Gd $5d$ states interacting with Cu $d$ and $s$ states and is mainly responsible for the bonding between Gd and Cu.\\

\subsubsection{Magnetic Moments}
The magnetic moment of Gd in hcp Gd within the muffin-tin sphere is $7.39~\mu_{\mathrm{B}}$. Since the muffin-tin sphere does not enclose the entire volume in the unit cell, we also consider the contribution from the interstitial region to the magnetic moment contributing $0.41~\mu_{\mathrm{B}}$ per atom. Adding the magnetic moment inside the muffin-tin and the interstitial region, the calculated total magnetic moment per atom in hcp Gd is $7.80~\mu_{\mathrm{B}}$, which is in good agreement with the experimental value of $7.63~\mu_{\mathrm{B}}$. As discussed by Kurz \textit{et al.}, a slight increment in the magnetic moment is due to small moment on $d$ electrons, which was calculated by us to be  $0.35~\mu_{\mathrm{B}}$. From the $l$-resolved magnetic moments, it is evident that the magnetic moment is mainly due to $4f$ electrons. 

As for GdCu, the magnetic moment is contributed mainly due to the spin imbalance within the muffin-tin sphere. Due to the anti-ferromagnetic configuration of the magnetic moments in GdCu, the spin density in the interstitial region integrate to zero magnetic moment and consequently does not contribute to the magnetic moment. The magnetic moment on Gd is $7.22~\mu_{\mathrm{B}}$, which is consistent with the experimental value of $7.24~\mu_{\mathrm{B}}$ by Blanco \textit{et al.}~\cite{blanco} Similar to the bulk Gd, we find a small spin-polarization on $d$ electrons due to the polarization of the $f$ electrons of Gd contributing $0.22~\mu_{\mathrm{B}}$ to the total magnetic moment of Gd. No induced spin polarization is observed on Cu $d$ states due to the AFM order of the Gd atoms. 

\begin{table*}[htp]
	\caption{Equilibrium lattice constants obtained for cubic GdCu within LDA+$U$ and GGA+$U$ density functionals, the comparison of lattice constant with FM and AFM configuration with respect to experimental lattice constant values ($\Delta a_{0}$), magnetic moments ($\lvert$M$\lvert$) in the muffin tin sphere. The last two big-columns show energy difference in meV/atom denoted by $\Delta E_A$, $\Delta E_C$ and $\Delta E_G$ between FM and three different AFM configurations A-, C- and G- types, respectively, as described in \eqref{deltaa}, \eqref{deltab} and \eqref{deltac}.  The calculated $J_{1}$, $J_{2}$ and $J_{3}$ are listed in the table for collinear and spin spiral calculations. }
\label{eqlc-fm-afm}	
	{\def\arraystretch{2}\tabcolsep=1pt
			\begin{tabular}{M{1.2cm}M{1.0cm}|M{0.9cm}M{0.9cm}|M{0.9cm}M{0.9cm}|M{0.9cm}M{1.0cm}M{0.9cm}|M{0.9cm}M{0.9cm}M{0.9cm}|M{0.9cm}M{1.0cm}M{0.9cm}|M{0.9cm}M{0.9cm}M{0.9cm}}
			\hline\hline
			&	&\multicolumn{2}{c}{FM} &\multicolumn{2}{c|}{AFM} &  \multicolumn{6}{c|}{Collinear} &  \multicolumn{6}{c}{Spin Spiral}  \\ \cline{3-4} \cline{5-6} \cline{7-12} \cline{13-18}
			& $a_{0}$(\AA) & $\Delta a_{0}$ (\%) & $M$ ($\mu_{\mathrm{B}}$)  & $\Delta a_{0}$ (\%) & $M$ ($\mu_{\mathrm{B}}$) & $\Delta E_A$ &   $\Delta E_C$ &   $\Delta E_G$ & $J_{1}$ & $J_{2}$  & $J_{3}$ &   $\Delta E_A$ &   $\Delta E_C$ &   $\Delta E_G$ & $J_{1}$ & $J_{2}$  & $J_{3}$ \\
			\hline\hline
			LDA+$U$ & $3.415 $& $-2.46$ & $7.18$  & $-2.46$ & $7.19$ &$1.14$ & $-11.27$& $13.10$ & $-0.04$ & $0.73$ & $-0.79$ & $1.35$ & $-11.31$ & $13.22$ & $-0.03$ & $0.72$ & $-0.80$ \\
			GGA+$U$ & $3.525$&$0.67$   &$7.18$  &$0.67$ & $7.22$ & $-4.01$ & $-22.56$ & $4.31$ & $0.89$ & $0.97$ & $-0.94$ & $-3.58$ &$-22.51$ & $4.76$ & $0.89$ & $0.96$ &$ -0.96$ \\
			\hline\hline
		\end{tabular}
	}
\end{table*}

We performed the self-consistent total-energy  calculations to determine equilibrium lattice constants for the FM and AFM ground states of bulk hcp Gd and cubic GdCu structures using LDA+$U$ and GGA+$U$ density functionals.  Our calculated value of the equilibrium lattice constant for GdCu using the LDA+$U$ functional is $3.415$~\AA, which underestimates the experimental value of $3.502$~\AA{}~\cite{iandelli,burzo,dongen} by $2.46\%$. On the other hand, the GGA+$U$ density functional gives $3.525$~\AA{} and is in better agreement with the experimental value (overestimating it by $0.67 \%$). Similar to the result at experimental lattice parameter, the AFM C-type  configuration is obtained as a  ground state of GdCu at equilibrium lattice constant. At equilibrium, the magnetic moment on the Gd atom within the antiferromagnetic configuration is 7.19 and 7.22~$\mu_{\mathrm{B}}$ for LDA+$U$ and GGA+$U$, respectively, both in good agreement with the experimental value of 7.24~$\mu_{\mathrm{B}}$ \cite{blanco}. The orbital resolved analysis depicts that Gd $4d$ electrons contribute up to $0.22~\mu_{\mathrm{B}}$ to the total magnetic moment. We have calculated the energy differences between FM and three AFM states at respective equilibrium lattice constant values and listed the results in Tab.~\ref{eqlc-fm-afm}. 

\subsection{Magnetic Interaction Parameters}
\label{sec:MIP}
\subsubsection{Collinear Magnetic States}
In order to describe thermodynamic properties of cubic GdCu, we develop a lattice spin model  and thus evaluate the exchange interaction parameters, $J_{ij}$, up to the third-nearest Gd neighbors by mapping the magnetic energy landscape of the system onto the classical Heisenberg Hamiltonian 
\begin{equation}
\mathcal{H}_{\mathrm{eff}}=\sum_{i\neq j}^{3} J_{ij}~ \mathbf{S}_{i} \cdot \mathbf{S}_{j}\, ,
\label{H-realspa}
\end{equation}  
where $J_{ij}$ is the exchange parameter between classical spins $\mathbf{S}_{i}$ and $\mathbf{S}_{j}$ (treated as vectors with the length $S_i=1$) of the magnetic atoms at different lattice sites $i$ and $j$. According to the choice of sign in \eqref{H-realspa}, $J_{ij}>0$ favors an antiferromagnetic coupling between a pair of spins. The mapping is realized by computing the energies for different collinear magnetic states, the FM state and three AFM configurations (\textit{viz.}, A, C, G) as shown in Fig.~\ref{3-antif} and compare those with Eq. \ref{H-realspa} restricting ourselves to three Gd neighbors. 

The energy differences relative to the FM state as listed in Tab.~\ref{eqlc-fm-afm} leads to a set of equations
\begin{eqnarray}
\Delta E_A= E_{\mathrm{A}}-E_{\mathrm{FM}}=&-4 J_{1} - 16 J_{2} - 16 J_{3} \label{deltaa}\\
\Delta E_C= E_{\mathrm{C}}-E_{\mathrm{FM}}=&-8 J_{1} - 16 J_{2} \label{deltab}\\
\Delta E_G = E_{\mathrm{G}}-E_{\mathrm{FM}}=&-12 J_{1} - 16 J_{3} \label{deltac}
\end{eqnarray}
 whose solution gives $J_{1}$, $J_{2}$, and $J_{3}$. The values are listed in Tab.~\ref{eqlc-fm-afm} for the LDA+$U$ and the GGA+$U$ methods. \\

\subsubsection{Spin-Spiral States}
An alternative approach to extracting above exchange parameters uses total-energy DFT calculations of noncollinear magnetic states described by a flat homogeneous spin-spiral state  elegantly realized in the density functional method using the generalized Bloch theorem~\cite{Sandratskii_1991}.  This procedure is very time-saving as it allows the calculation of the magnetic structure for an arbitrary spin-spiral vector $\mathbf{q}$ on the basis of the chemical, \textit{i.e.}, CsCl, unit cell.  One additional value of homogeneous spin spirals lies in the observation that they are also solutions of the classical Heisenberg model for periodic lattices. 

By virtue of a periodicity of GdCu, it is convenient to replace the quantities in \eqref{H-realspa} by their Fourier transformed equivalents. By exploiting the translational invariance of the lattice, we can write
\begin{equation}
\mathcal{H}_{\mathrm{eff}}= -N \sum_{\mathbf{q}} J({\mathbf{q}}) ~ S_{\mathbf{q}}\cdot S_{\mathbf{-q}} \, ,
\label{hei-fourier}\end{equation}
where $J({\mathbf{q}})$ represents lattice Fourier transform of the exchange interaction and is given by 
\begin{equation}
J(\mathbf{q})=\sum_{R} J_{\mathbf{0R}} \exp(i \mathbf{q\cdot R})\, .
\end{equation}
Here, $\mathbf{q}$ denotes a propagation vector of the spin spiral and $\mathbf{R}$ represents atom sites. The related flat spiral magnetic structure, characterized by a single wavevector $\mathbf{q}$ (single-q state), is defined by the Cartesian coordinates of the magnetization vector $\mathbf{M}_{i}$ given by
\begin{equation}
\mathbf{M}_{i} = M\big[\cos(\mathbf{q}\cdot\mathbf{R}_{i}),~ \sin(\mathbf{q}\cdot\mathbf{R}_{i}),~ 0\big]\, ,
\label{spir-e}\end{equation}
where $M$ is the size of the magnetic moment. In this context the above discussed magnetic configurations FM, G, A, and C  are equivalent to spin-spiral states at high-symmetry points, $\Gamma$, $R$, $X$, $M$,  of the cubic Brillouin zone with  propagation vectors of $\mathbf{q}=\frac{2\pi}{a}(0,0,0)$, $\mathbf{q}=\frac{2\pi}{a}(\frac{1}{2},\frac{1}{2} , \frac{1}{2})$, $\mathbf{q}=\frac{2\pi}{a}(\frac{1}{2},0, 0)$ and $\mathbf{q}=\frac{2\pi}{a}(\frac{1}{2},\frac{1}{2} , 0)$, respectively.

\begin{figure}
\includegraphics[scale=0.35, angle=270]{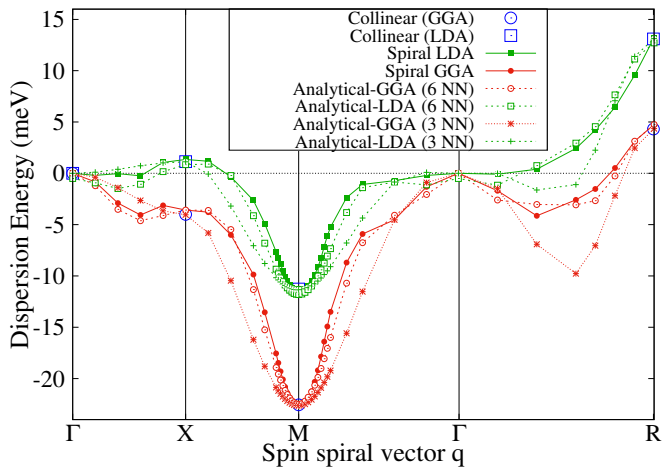}
\caption{The total energy $E(\mathbf{q})$ as function of the spin spiral $\mathbf{q}$-vector, calculated within GGA+$U$ (shown in filled circles) and LDA+$U$ (shown in filled squares). The total energy at high symmetry points calculated using collinear calculations for GGA+$U$ and LDA+$U$, are marked by open circles and open squares, respectively for comparison. The curves with the dotted lines represent the values obtained using analytical expressions of J($\mathbf{q}$) as in \eqref{analy-q}, considering both $J_{1},\dots ,J_{3}$ (shown as 3 NN) and $J_{1},\dots ,J_{6}$ (shown as 6 NN)  for GGA+$U$ (red) and LDA+$U$ (green). } 
\label{sps}
\end{figure}

The total energy difference $\Delta E(\mathbf{q})$ relative to the FM state is computed self-consistently as a function of spin-spiral vector $\mathbf{q}$ on a fine $q$-grid along the high-symmetry lines of the cubic Brillouin zone. In order to resolve the energy with sufficient accuracy the k-point integration was performed  on $24\times24\times24$  regular $k$-point mesh. The results are shown in Fig.~\ref{sps} for cubic GdCu. It is revealed from Fig.~\ref{sps} that the spin spirals have a narrow energy minimum at the high-symmetry $M$-point,  which corresponds to the C-type AFM configuration, as shown in Fig.~\ref{3-antif}. This is in agreement with the results of the collinear calculations discussed early this section. 

Overall there is an excellent agreement between the energy differences at the high-symmetry points calculated by the spin-spiral approach and the collinear calculations as can be seen in Tab.~\ref{eqlc-fm-afm} underscoring the reliability of our calculations. For example,
the energy gain with respect to FM  due to the spin-spiral state at the $M$-point obtained with the GGA+$U$ functional is $22.51$~meV/atom, which is in excellent agreement with that of $22.56$~meV/atom obtained through the collinear calculation.  The LDA+$U$ calculation exhibits a smaller energy gain of $11.31$~meV/atom in agreement with its collinear counterpart. 

 Although the qualitative behavior of the energy dispersion of the spin-spiral is the same as function of $\mathbf{q}$ for LDA+$U$ and GGA+$U$, we note a significant energy shift between the two, whose magnitude depends on the wave vector $\mathbf{q}$ and  thus on the magnetic structure. This can even lead to sign changes. For example, at the $X$-point, the dispersion energy has positive (1.35~meV/atom) and negative values ($-$3.58~meV/atom) for LDA+$U$ and GGA+$U$ functionals, respectively, although small in magnitude. This further indicates that the nearest neighbor Heisenberg interaction, $J_1$, are less dominated than the next-nearest neighbor interactions,  $J_2$, consistent with the value tabulated in Tab.~\ref{eqlc-fm-afm}. Interestingly, the $R$-point equivalent to G-type AFM configuration (Fig.~\ref{3-antif}) resides at an energy higher than the $\Gamma$-point, irrespective of the choice of GGA+$U$ or LDA+$U$ functional, in agreement with the respective collinear results. The size of the energy shift depends on the lattice parameters. Figure ~\ref{sps} presents the results  for the respective calculated equilibrium parameters.  Taking the same lattice parameters for GGA+$U$ or LDA+$U$, the  quantitative behavior is much closer (not shown in the Figure). \\

We compare the calculated values of  $J_{1}$, $J_{2}$ and $J_{3}$ using collinear calculations (real space Hamiltonian as in Eq. \eqref{hei-fourier}) and using spin spiral calculations (as in Eq. \eqref{hei-fourier}) listed in Tab.~\ref{eqlc-fm-afm}. If we fit the three exchange parameters to the three energy differences at the high-energy points $M$, $K$, and $R$ we obtain practically the same parameters due to the excellent quantitative agreement between the energy differences obtained by the two different approaches.  It can be observed from Tab.~\ref{eqlc-fm-afm} that in case of the GGA+$U$ results the magnitudes of $J_{1}$, $J_{2}$ and $J_{3}$ are closely equal to $1$ (except $J_1$ for LDA+$U$). For GGA+$U$, both $J_{1}$ and $J_2$ are positive and favor antiferromagnetic coupling. $J_{1} > 0$ favors a G-type structure (Fig.~\ref{3-antif}), where all nearest neighbors (NN) have an antiferromagnetic configuration. $J_{2} > 0$ means, all next-nearest neighbors (NNN's) \textit{i.e.}\ all face-diagonal atoms of the cube interact antiferromagetically. Since $|J_{1}|$ and $|J_{2}|$ are almost same in magnitude, their competition leads to spin frustration. In addition, third-nearest neighbors, \textit{i.e.}\ atoms across the cube diagonal interact ferromagnetically ($J_{3} < 0$), which results in planes of  square lattices with antiferromagnetic configurations coupling ferromagnetically from plane to plane favoring C-type antiferromagnetic order (Fig.~\ref{3-antif}). For LDA+$U$, the next-nearest neighbor interactions are clearly dominant leading to the C-type configuration. \\

\subsubsection{Multi-q State}
\begin{figure}[b]
\includegraphics[width=3.5in, angle=0]{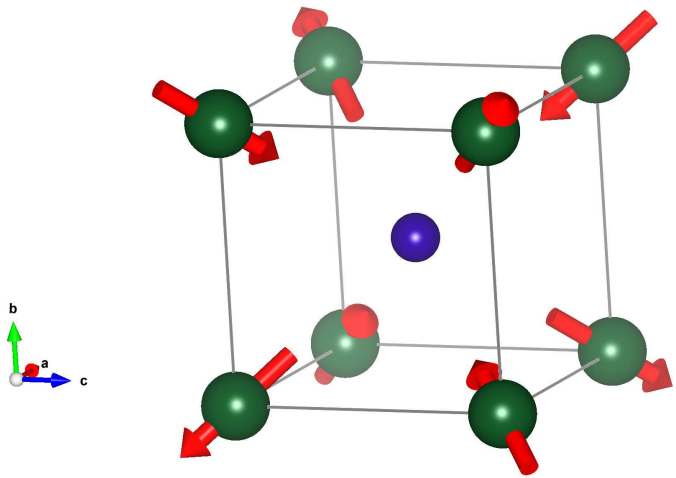}
\caption{(color online) An image of the 3$\mathbf{Q}$ structure for the cubic GdCu compound, with spins pointing in all three diagonal directions of the cube. Big green (small blue) sphere symbolizes the Gd (Cu) atoms. The direction of the magnetic moment of a corner atom and its face-diagonal neighbor are related by  a $\approx 109.47^\circ$ (tetrahedral angle) rotation around a space-diagonal rotation axis.}
\label{multiq}
\end{figure}
By virtue of the rotational invariance of the Heisenberg exchange interaction energy (excluding spin-orbit interaction) with respect to the underlying lattice, magnetic states described by symmetry equivalent $\mathbf{q}$-vectors have the same Heisenberg energy, as in Eq. \eqref{hei-fourier}. For example, wave vector $ \mathbf{q}$ and the time reverse state $-\mathbf{q}$ have the same energy and both form a single-$q$  spin-spiral state. High-symmetry points in the Brillouin zone have a multiple of Heisenberg degenerate $\mathbf{Q}$-vectors, whose superposition can form a multi-q state, when the spins retain their lengths, this is a much more complex magnetic structure that exhibits, on the level of the Heisenberg interaction, the same energy as the single-q state. Higher-order magnetic interactions beyond the Heisenberg model contained in the interactions described by DFT, can then lead to deviations to the single-q state, \textit{i.e.}\  a multi-q state can be higher or lower in energy than the single-q one. 

We found that the  lowest energy configuration for cubic GdCu compound is C-type AFM, a single-q structure, which corresponds to the $M$-point in the three-dimensional Brillouin zone.  The Brillouin zone of the three-dimensional lattice contains three symmetry-equivalent $M$-points,  denoted as $\mathbf{Q}_{(k)}$, for $k=1, 2, 3$, with $\mathbf{Q}_{1}=\pm\frac{2\pi}{a}(\frac{1}{2},0, \frac{1}{2})$, $\mathbf{Q}_{2}=\pm\frac{2\pi}{a}(0, \frac{1}{2},\frac{1}{2})$,  and $\mathbf{Q}_{3}=\pm\frac{2\pi}{a}(\frac{1}{2},\frac{1}{2}, 0)$. The orthonormalized linear combination of the three spin-spirals with wave vectors $\mathbf{Q}_{(k)}$ can then from a $3\mathbf{Q}$ state,\cite{satoru} which is a non-collinear structure as shown in Fig.~\ref{multiq} and represented by
 \begin{eqnarray}
  M_i =
  \frac{M}{\sqrt{3}}[\cos(\mathbf{Q}_{1}\cdot\mathbf{R}_{i}) ,\cos(\mathbf{Q}_{2}\cdot\mathbf{R}_{i}) ,\cos(\mathbf{Q}_{3}\cdot\mathbf{R}_{i})]\, .
  \end{eqnarray}
  
  Employing a supercell containing 8 chemical unit cells, imposing the C-type antiferromagnetic state as well as  the $3\mathbf{Q}$ state, we determined self-consistently the total energy difference between the single- and triple-$q$ state. Employing the GGA-$U$ functional, we found that the $3\mathbf{Q}$-state is 3.80~meV/atom higher in energy than the C-type AFM order, confirming the absence of the non-collinear magnetic ground state. In principle, also other high-symmetry q-points can form multi-q states, but their single-q states are so much higher in energy than the C-type AFM one, so that their superpositions are improbable to become the ground state. 

\subsection{The N\'{e}el temperature \label{sps-neel}}
Since the Gd-$4f$ magnetic moment is rather localized and thus its magnitude depends little on the relative orientation to neighboring moments, the Heisenberg Hamiltonian is a good approximation to estimate the N\'{e}el temperature. Here we employ two approaches: the mean field approach (MFA) and the random-phase approximation (RPA). Within MFA, the N\'{e}el temperature ($T_{\mathrm{N}}$) of the spin spiral with wave vector $\mathbf{Q}$  is given according to  Ref.~\onlinecite{turek, turek2,kubler} as
\begin{equation}
k_{\mathrm{B}} T_{\mathrm{N}}^{\mathrm{MFA}} = \frac{2}{3} J(\mathbf{Q})\, ,
\end{equation}
where $k_{\mathrm{B}}$ is the Boltzmann constant. Considering the choice of sign  of Hamiltonian \eqref{H-realspa}, $J(\mathbf{Q})$ is the absolute maximum of $J(\mathbf{q})$ scanned over the entire Brillouin zone obtained at $\mathbf{q}=\mathbf{Q}$ and  the maximum of $J(\mathbf{Q})$ corresponds to minimum of  the energy of the single-q mode (for details, see Ref. \cite{jensen, long, yamamoto,benedikt}) according to 
\begin{equation}
E=-N S^{2} J({\mathbf{Q}})\, .
\label{ene}\end{equation}

Taking into account that the calculated minimum of Eq. \eqref{ene} is at the $M$-point ($\mathbf{Q}=\frac{2\pi}{a}(\frac{1}{2},\frac{1}{2} , 0)$), \textit{cf.}\ Fig.~\ref{sps}, the calculated N\'eel temperature within GGA+$U$ is equal to  $T_{\mathrm{N}}^{\mathrm{MFA}}=174.53$ K. This value is overestimated by $16$~\%  with respect to the experimental value of $T_{\mathrm{N}}^{\mathrm{Expt}}=150$~K. Using the LDA+$U$ functional, a value of $T_{\mathrm{N}}^{\mathrm{MFA}}=87.19$~K was obtained,  $42$~\% lower than the experimental value. It is a well-known fact~\cite{turek, turek2,kubler} that, MFA overestimates the critical temperature, for simple cubic magnetic lattices even more than of compact lattices. Therefore, we can conclude the GGA+$U$ approximation to the exchange correlation energy functional gives a much better description of the magnetic exchange interaction than the LDA+$U$ functional. Again this is mostly an effect of GGA+$U$ lattice parameter. \\

An improved estimation for the  $T_{\mathrm{N}}$ is provided by the random phase approximation (RPA)\cite{turek,tyablikov}, since RPA weights low-energy excitations with wave vectors $\mathbf{q}$ in the vicinity of the $M$-point,  $\mathbf{Q}$, by the inverse power. It is given by
\begin{equation}
\frac{1}{k_{\mathrm{B}} T_{\mathrm{N}}^{\mathrm{RPA}}}=\frac{3}{4} \frac{1}{N}\sum_{\mathbf{q}} \left\lbrace\frac{1}{\left[J(\mathbf{Q})-J(\mathbf{q}) \right]} + \frac{1}{\left[ W(\mathbf{q},\mathbf{Q})\right]} \right\rbrace \, ,
\end{equation}
where $N$ denotes number of $\mathbf{q}$ vectors considered, and 
\begin{equation}
W(\mathbf{q},\mathbf{Q})=J(\mathbf{Q})-\frac{1}{2}J(\mathbf{q}+\mathbf{Q})-\frac{1}{2}J(\mathbf{q}-\mathbf{Q})\, .
\end{equation}

In order to compute the $T_{\mathrm{N}}$ within RPA, a very good approximation of  the total energy E($\mathbf{q}$) by the respective $J(\mathbf{q})$ is needed for a dense mesh of $\mathbf{q}$ vectors  throughout the Brillouin zone. This cannot be achieved with the parameters $J_1, \dots , J_3$ discussed in section~\ref{sec:MIP}  involving the interaction between only three neighbors.  In order to reduce the computational cost of calculating the total energy  $E(\mathbf{q})$ on a dense $\mathbf{q}$-grid, we computed $E(\mathbf{q})$ on a dense grid along high-symmetry lines as shown in Fig.~\ref{sps}, and reproduced the results using exact analytical expressions as shown in appendix \ref{append}, using exchange constants fitted up to sixth-nearest neighbors.  Using this analytical expression, we reproduce  the spin-spiral total energy results as shown in Fig.~\ref{sps} by dotted green and red lines to an excellent degree.  Using this approach, we estimate the N\'eel temperature, $T^{\mathrm{RPA}}_{\mathrm{N}}$, within RPA for GGA+$U$ and LDA+$U$ methods to $122.9$~K and $73.7$~K,  which  is $0.77$ and $0.84$ times that of MFA values, respectively.  This is reasonable as it is known that for the nearest-neighbor approximation to the Heisenberg exchange parameter the $T_{\mathrm{N}}$ values calculated within RPA are only 66~\% of the MFA value for simple cubic structure.\cite{tyablikov}

Since RPA is a very good approximation in particular for large-moment systems like GdCu coming close to the Monte Carlo values for the N\'eel temperature we conclude the Curie temperature of $122.9$~K as calculated in GGA+$U$ is underestimated by 25~\% with respect to the experimental value of 
$T_{\mathrm{N}}^{\mathrm{Expt}}=150$~K. As already seen for the MFA, in LDA+$U$ the N\'{e}el temperature is not sufficiently well reproduced. This is mainly an effect of the equilibrium lattice constant,  experimental value  of which is less well represented by  LDA+$U$ than by  GGA+$U$, and the fact that the exchange parameter $J$ depend significantly at the lattice constant.

\subsection{Effective Coulomb interaction and core states in GdCu\label{core}}

\subsubsection{cRPA method}\label{crpa-des}
\begin{figure}[t]
	\includegraphics[scale=0.34, angle=0]{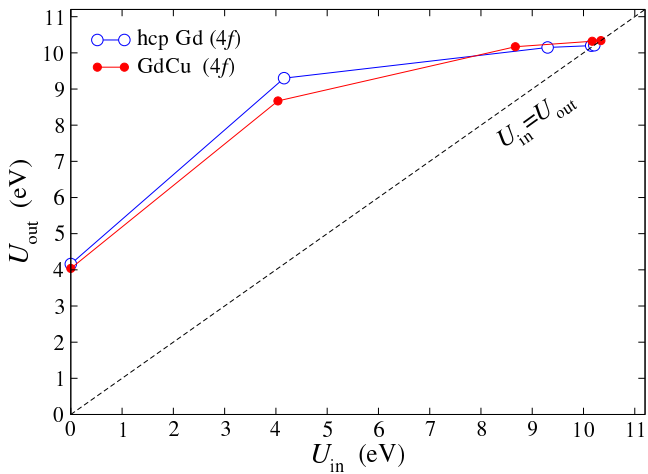}
	\caption{Self-consistent calculation of effective Coulomb interaction parameter (Hubbard $U$) between
		localized $4f$ electrons in hcp Gd and GdCu in CsCl-type structure.}
	\label{rpa-u}
\end{figure}

In this subsection, we discuss the strength of the effective Coulomb interaction (Hubbard $U$) between the localized 
$4f$ electrons and theoretical understanding of the spectra of GdCu core levels. We calculated the Hubbard $U$ 
parameter for hcp Gd and GdCu using self-consistent cRPA method at experimental lattice parameter.
By employing the self-consistent constrained random-phase approximation (cRPA) \cite{cRPA_a, cRPA_b,cRPA_1,cRPA_1a,cRPA_2,cRPA_3, cRPA_3a} within 
the \texttt{SPEX} code\cite{Spex,Wannier90} we calculate the strength of the effective Coulomb interaction (Hubbard $U$) 
between localized $4f$ electrons in hcp Gd and cubic GdCu (for further technical details see Refs.\,\onlinecite{cRPA_Sasioglu} 
and \onlinecite{Sasioglu}). We use a $8 \times 8 \times 5$  and  $3 \times 3 \times 5$  $\mathbf{k}$-point grid
for hcp Gd and GdCu (with C-type AFM order in tetragonal unit cell), respectively in the cRPA calculations. The cRPA $U$ values are turned out to be large, \textit{i.e.},  
$U=10.21$ eV for hcp Gd and $U=10.34$ eV for GdCu  and thus we use a smaller $U$ value in LDA+$U$ and GGA+$U$ calculations. 
 To calculate 
$U$ self consistently we start with a standard GGA calculation as an input for cRPA method in the Spex code and then obtain 
the initial $U$ parameter to be used in GGA+$U$ calculation.  Then the procedure is repeated till the self-consistency is 
reached, i.e.,  $U_{\textrm{out}}= U_{\textrm{in}}$. The obtained results are presented in Fig.~\ref{rpa-u}. As seen 
DFT-GGA gives $U$ values just above 4~eV for both materials and the self-consistent calculations converge in few steps. 
The final converged values for Hubbard $U$ parameter for hcp Gd and GdCu are $10.21$ and $10.34$~eV, respectively. Our
Hubbard $U$ parameter for hcp Gd is in good agreement with previous calculations \cite{U_Karlsson,U_Sakuma}. However, it
is known that the RPA method (cRPA as well) overestimates the Coulomb interaction for localized orbitals, especially for  
$4f$ systems, in which exchange splitting of the $4f$ states turns out to be too large compared to experiments. E.g., in 
quasi-particle self-consistent $GW$ (QSGW) calculations, such a large splitting is attributed to the overestimation of 
the strength of the  screened Coulomb interaction in QSGW method, which stems from the  neglect of interaction between  
electron-hole pairs in its intermediate states (excitonic effects)~\cite{4f_GW}. Due to this, the band gaps are too large in 
semiconductors within QSGW method and this gap overestimation systematically increases with localization of the 
orbitals~\cite{SC_GW}.

\subsubsection{DFT+$U$ method}
\begin{figure}[b]
\includegraphics[scale=0.35, angle=270]{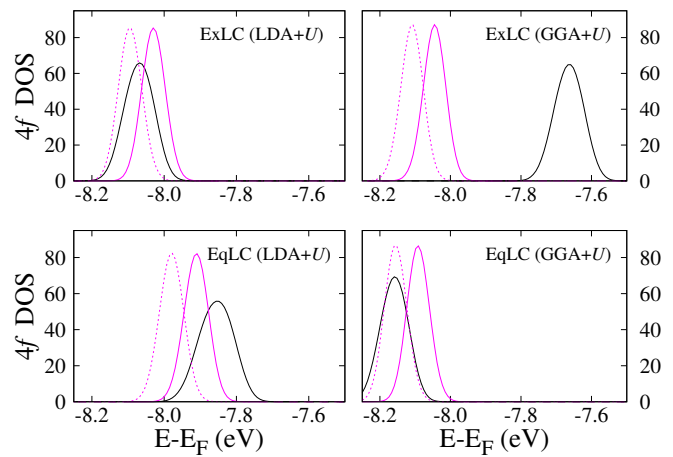}
\caption{$4f$ states of hcp Gd and GdCu using LDA+$U$ and GGA+$U$ methods at experimental lattice constant (ExLC) and calculated equilibrium lattice constant (EqLC).  $E_{\mathrm{F}}$ on X-axis is set to zero. The $4f$ states for Gd (with $U=6.7$~eV) is shown in black curve, whereas in GdCu $4f$ are shown in magneta color,  with $U=6.70$~eV and $U=6.83$~eV indicated by continuous and dotted lines, respectively.}
\label{4f-dft}
\end{figure}

Since the self-consistent cRPA method overestimates the $U$ parameter in $4f$ materials, we use the soft Hubbard $U$ parameter 
for hcp Gd  as calculated by Shick \textit{et al.} \cite{shick} and used by Kurz \textit{et al.} \cite{kurz} 
The $U$ parameter in GdCu is chosen in two ways: (a) same as for bulk Gd, (b) an additional shift in $U$ parameter($\Delta~U$) of $0.13$~eV  with respect to bulk Gd, as suggested by cRPA method in Sec. \ref{crpa-des}. 
The result for unchanged $U$ (i.e., $U=6.7$~eV) as 
well as modified $U$ (i.e., $U=6.83$~eV) for GdCu are shown in Fig.~\ref{4f-dft}. We compare our results with the experimentally observed shifting of $4f$ peak in GdCu with respect to bulk Gd of 0.3 eV below $E_{\mathrm{F}}$.\cite{szade1, szade2,lachnitt}

First, we discuss the results calculated at the experimental lattice parameters of Gd and GdCu. The $4f$ peak in hcp Gd is produced  
at $8.1$~eV  below E$_\mathrm{F}$ using LDA+$U$ method. However, the positive  shift relative to 
$E_\mathrm{F}$ is too small in GdCu using $6.83$ eV for the $U$  parameter. Instead, if $U$ parameter is $6.7$ eV in 
GdCu and hcp Gd, $4f$ is observed to shift in opposite direction than that observed in the experiment. A positive shift in 
agreement with the experiment is observed at the experimental lattice constant if GGA+$U$ method is used, however exact 
location of $4f$ is not produced for both choices of $U$.  

For completeness, we present results of $4f$ peaks computed at equilibrium lattice parameter (EqLC). Within LDA+$U$, a small positive shift ($0.1$~eV) is observed in GdCu with respect to hcp Gd, at EqLC, however, the location of $4f$ peaks are underestimated as compared to the experimental observation. The GGA+$U$ method produces approximately correct location of the $4f$ peak ($8.1$ eV) in hcp Gd at EqLC. However, an opposite shift is observed with respect to the experimental finding, and is shown in Fig.~\ref{4f-dft}. In addition to the consideration of $U$ parameter for $4f$ in Gd, we also considered $U=2.91$~eV and $J=1.26$~eV for Cu $3d$ states, however correct $4f$ shift in GdCu in relative to hcp Gd, as in the experiments is also not observed. \\

\subsubsection{Slater - Janak Transition state theory}
\begin{figure}
\includegraphics[scale=0.35, angle=270]{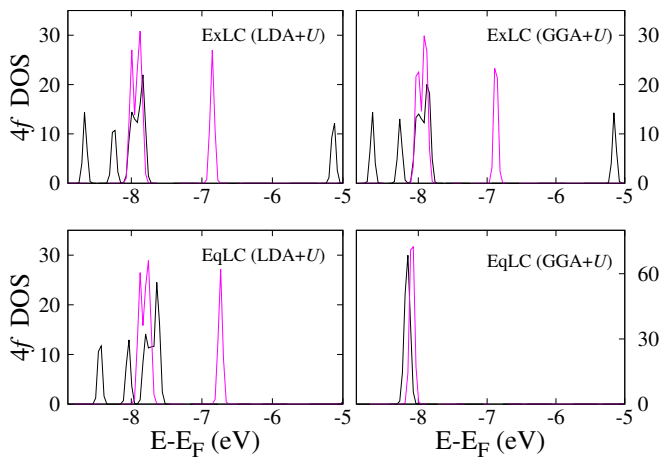}
\caption{$4f$ states of hcp Gd and GdCu using LDA+$U$ and GGA+$U$ methods at experimental lattice constant (ExLC) and equilibrium lattice constant (EqLC) calculated using Slater-Janak transition state theory.  $E_{\mathrm{F}}$ on X-axis is set to zero. The $4f$ states for Gd is shown in black line, whereas magneta colored lines shows GdCu $4f$ peaks.}
\label{4f-sj}
\end{figure}

Further we consider the details of XPS experiments used to investigate the core level shifts. The XPS binding energy of the core level is achieved in the experiments by ejection of a core electron to the infinity under X-ray irradiation. This ejected electron creates positively charged core hole and is screened by other electrons in a system. This can be evaluated by extending DFT based on Slater-Janak transition-state approach,\cite{janak1,slater-1974} in which eigenenergy is obtained by considering half occupation of the orbital of interest and placed into the valence band. Our calculations using this approach were performed by removing half an electron from the $j=\frac{5}{2} ~ 4f$ state. Figure~\ref{4f-sj} shows the $4f$ peaks calculated at $U=6.7$~eV for hcp Gd and bulk GdCu using Slater-Janak transition state approach. It can be seen that, positive shift of $4f$ levels in GdCu relative to hcp Gd is not observed irrespective of the lattice parameter and the exchange functional. \\

\subsubsection{Hubbard-I approximation}
To investigate in more detail about the $4f$ shift in GdCu, we performed calculations in the LDA + Hubbard-I (LDA+HIA) approximation 
(with the crystal field and SOC included) at the experimental lattice parameter.
Details of the implementation used in this work are given 
elsewhere\cite{hia}, and we refer the reader to this paper for
a complete description of our computational method.
The calculations were performed making use of two types of the double counting (DC): the ground mean field limit (AMF) and fully localized limit (FLL) for (i) ferromagnetic hcp-Gd, and (ii) anti-ferromagnetic (type C) GdCu. \\

In the Tab.~\ref{table-hub},  the spin ($M_S^f$), orbital ($M_L^f$) magnetic moments for the Gd atom $f$ -shell are listed, together with the spin moment for the 5$d$-electrons, and the total spin moment $M_S^{TOT}$ per formula unit. It is seen that the moments are almost independent on the choice of the DC. 

\begin{figure}[!htbp]
\centerline{\includegraphics[scale=0.33, angle=270]{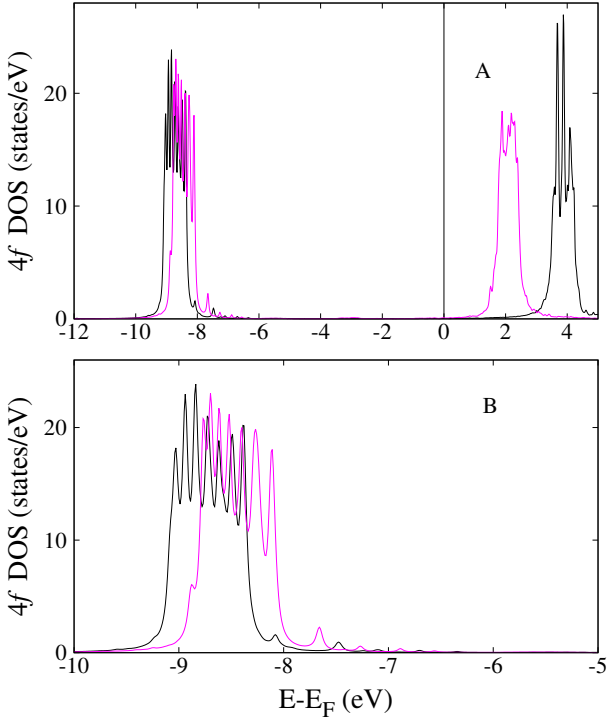}}
\caption{4$f$ states ferromagnetic hcp Gd (black) and  anti-ferromagnetic (type C) GdCu (magenta) calculated making use of LDA+HIA with FLL-DC.
(A) full energy interval, (B) narrower energy interval near binding energy.}
 \label{fig3-sasha}
\end{figure}

In the Fig.~\ref{fig3-sasha},  we show the $f$-DOS for Gd in ferromagnetic hcp Gd versus GdCu in C-type AFM. It is seen  there is a small negative shift of the binding energy of $f$-Gd in GdCu with respect to the hcp-Gd. This is similar to  the DFT+$U$ results shown at the beginning of this section, and contradicts to the experimental findings. Note that the binding energy shift is very similar  
for both AMF-DC and FLL-DC.

In short, we remark that, the experimental observation of $4f$ core level shift in GdCu with respect to bulk Gd is not reproduced using $ab ~initio$ method as well as Hubbard-I approximation consistently. We encourage more experiments to strengthen the arguments regarding the observed $4f$ shift in GdCu compound. \\

\begin{table}[htbp]
\caption{Spin and orbital magnetic moments in $\mu_\mathrm{B}$, calculated
for ferromagnetic hcp-Gd, and anti-ferromagnetic (type C) GdCu}
\setlength{\tabcolsep}{12pt}
\renewcommand{\arraystretch}{1.4}
\begin{tabular}{ccccccc} 
\hline\hline
&\multicolumn{4}{c}{hcp-Gd}\\\cline{2-5}
 
& $M_S^f$ & $M_L^f$ & $M_S^d$ & $M_S^{TOT}$ \\
\hline
AMF&6.87  & 0.06  & 0.41  & 7.70  \\
FLL&6.90  & 0.04  & 0.41  & 7.74 \\
\hline

&\multicolumn{4}{c}{GdCu}\\\cline{2-5}
& $M_S^f$ & $M_L^f$ & $M_S^d$ & $M_S^{TOT}$ \\
\hline
AMF&6.70  & 0.13  & 0.26  & 0  \\
FLL&6.81  & 0.07  & 0.25  & 0 \\
\hline
\end{tabular}
\label{table-hub}
\end{table}


\section{Conclusions\label{con}}

We investigated the cubic GdCu (CsCl-type) compound, treating the $4f$ states within LDA+$U$ 
and GGA+$U$ approximation to the exchange-correlation functional. The 
structural parameters of GdCu within GGA+$U$ agree well with the experimental results. Based 
on collinear as well as non-collinear magnetic calculations, we found that GdCu settles in the C-type antiferromagnetic  order  at low temperatures, a magnetic structure describable by a flat magnetic spin-spiral state with a wave 
vector  $\mathbf{Q}=\frac{2\pi}{a}(\frac{1}{2},\frac{1}{2} , 0)$, consistent with the experiments. 
An associated triple-q state was found to have a higher energy. The calculated value of N\'{e}el temperature of GdCu using the GGA+$U$ approximation  amounts to $ 174.5$~K
and $122.9$~K, in the mean field and random phase approximation, respectively, which is in reasonable agreement to the experimental value of $150$~K. To describe the energy landscape with sufficient accuracy a Heisenberg model is required that includes exchange parameters  at least up to the sixth-nearest neighbor interaction. The equilibrium lattice constant obtained within the LDA+$U$ approach is 2.5~\% smaller than the experimental one. This is sufficient to substantially change the Heisenberg exchange parameter to the point that they cannot describe reliably the magnetic interactions of GdCu, although the ground state is still C type AFM. 

Using the calculated shift of Hubbard $U$ 
value between hcp Gd and cubid GdCu, as obtained by the constrained random phase approximation, we found that the $4f$ shift 
in GdCu with respect to hcp Gd is not consistent with the experiments. The calculations 
performed using Hubbard-1 approximations are in agreement with our DFT+$U$ results indicating the necessity of the experimental investigations 
in more detail.

\begin{acknowledgments}
V. K. acknowledges DST-SERB overseas postdoctoral fellowship during the course of this work. We thank fruitful discussions with Dr.\ Daniel Wortmann, Dr.\ Gregor Michalicek  and Jens Br\"{o}der for this work. We also gratefully acknowledge the  J\"ulich Supercomputing Centre and RWTH Aachen University for providing computational resources under projects jara0161, jiff40 and cias-1. 
A.B.S. acknowledges financial support provided by OPVVV project SOLID21 - CZ.02.1.01/0.0/0.0/16$_{-}$019/0000760, 
and by the GACR grant 18-06240S. 
\end{acknowledgments}

\appendix
\section{Analytical expression for $J(\mathbf{q})$}\label{append}

The energetics of magnetic states on simple cubic lattice is described within the model Hamiltonian using the Fourier transform of the exchange constants $J\mathbf{(q)}$ as given in equation \eqref{hei-fourier}. The $\mathbf{q}$ is expanded in terms of primitive vectors of the reciprocal lattice, $\mathbf{q}=q_{1}\mathbf{b_{1}}+q_{2}\mathbf{b_{2}}+q_{3}\mathbf{b_{3}}$. The exchange interaction is considered up to sixth nearest neighbor, $J\mathbf{(q)}$ is expressed as

\begin{small}
\begin{eqnarray*}
J(\mathbf{q})&=&2 J_{1} \bigl[ \cos(2\pi q_{1})+\cos(2\pi q_{2}) + \cos(2\pi q_{3})\bigr]   \nonumber \\
& +& 2 J_{2} \bigl[ \cos(2\pi (q_{1}+q_{2}))+\cos(2\pi (q_{1}+q_{3})) + \cos(2\pi (q_{2}+q_{3}))  \nonumber \\
& & +  \cos(2\pi (q_{1}-q_{2}))+\cos(2\pi (q_{1}-q_{3})) + \cos(2\pi (q_{2}-q_{3})) \bigr] \nonumber \\
& +& 2 J_{3} \bigl[ \cos(2\pi (q_{1}+q_{2}+q_{3}))+ \cos(2\pi (q_{1}-q_{2}+q_{3}))  \nonumber \\
& & +  \cos(2\pi (q_{1}+q_{2}-q_{3}))+\cos(2\pi (q_{1}-q_{2}-q_{3})) \bigr]  \nonumber \\
&+ &  2 J_{4} \bigl[ \cos(2\pi (2 q_{1}))+ \cos(2\pi (2 q_{2})) + \cos(2\pi (2 q_{3}))\bigr]  \nonumber \\
&+ &  2 J_{5} \bigl[\cos (2\pi(2q_{1}+q_{2}))+\cos (2\pi(2q_{1}+q_{3}))+\cos (2\pi(2q_{2}+q_{3})) \nonumber \\
& & +  \cos (2\pi(q_{2}+2q_{3}))+\cos (2\pi(q_{1}+2q_{3}))+\cos (2\pi(q_{1}+2q_{2})) \nonumber \\
& & +  \cos (2\pi(2q_{1}-q_{2}))+\cos (2\pi(2q_{1}-q_{3}))+\cos (2\pi(2q_{2}-q_{3})) \nonumber \\
& & +  \cos (2\pi(q_{2}-2q_{3}))+\cos (2\pi(q_{1}-2q_{3}))+\cos (2\pi(q_{1}-2q_{2}))\bigr]  \nonumber \\
& + & 2 J_{6} \bigl[ \cos(2\pi(q_{1}+q_{2}+2q_{3}))+ \cos(2\pi(q_{1}+2q_{2}+q_{3}))\nonumber\\
& &+  \cos(2\pi(2q_{1}+q_{2}+q_{3}))+ \cos(2\pi(q_{1}+q_{2}-2q_{3}))\nonumber\\
& & +  \cos(2\pi(q_{1}-2q_{2}+q_{3}))+ \cos(2\pi(-2q_{1}+q_{2}+q_{3}))\nonumber\\
& & +  \cos(2\pi(q_{1}-q_{2}-2q_{3}))+ \cos(2\pi(q_{1}-2q_{2}-q_{3}))\nonumber\\
& & +  \cos(2\pi(-2q_{1}+q_{2}-q_{3}))+ \cos(2\pi(-q_{1}+q_{2}-2q_{3}))\nonumber \\
& & +  \cos(2\pi(-q_{1}-2q_{2}+q_{3}))+ \cos(2\pi(-2q_{1}-q_{2}+q_{3}))\bigr] \\
\\ \label{analy-q}\end{eqnarray*}
\end{small}

The $J_{1}$,$J_{2}$, ..., $J_{6}$ in \eqref{analy-q} are obtained by calculating spin spirals at $q=\frac{2\pi}{a}(\frac{1}{2},0,0)$,  $q=\frac{2\pi}{a}(\frac{1}{2},\frac{1}{2},0)$,  $q=\frac{2\pi}{a}(\frac{1}{2},\frac{1}{2},\frac{1}{2})$,  $q=\frac{2\pi}{a}(\frac{1}{2},\frac{1}{4},0)$, $q=\frac{2\pi}{a}(\frac{1}{4},\frac{1}{4},0)$  and  $q=\frac{2\pi}{a}(\frac{1}{4},\frac{1}{4},\frac{1}{4})$ within LDA+$U$ and GGA+$U$ methods as in Sec. \ref{sps-neel}. The obtained values are listed in Tab.~\ref{j1j6list}.

If we compare the exchange parameters $J$ in Tab.~\ref{eqlc-fm-afm} and Tab.~\ref{j1j6list}, we notice that the value $J_4$ describing the interaction between the fourth-nearest neighbor is still rather large, but missing in the discussion above. To illustrate, we plotted the energy landscape of $J(\mathbf{q})$ in Fig. \ref{sps}, using the analytic expression \ref{analy-q}, but with $J_{1}, \dots ,J_{3}$ as well as with $J_{1}, \dots ,J_{6}$. It can be observed from Fig. \ref{sps} that, with the parameters $J_1, \dots ,J_3$  we are able to parameterize the energy landscape related to the magnetic states in Fig.~\ref{3-antif}, but this is not sufficient to describe the energy landscape on level to study dynamical and thermodynamical properties. One finds that by inclusion of more interaction parameters, the values of $J_1$ and $J_2$, change nearly be a factor two, at least for the values obtained within GGA+$U$. The improved energy landscape by including more $J$'s in evaluating $J(\mathbf{q})$ improves $T_{\mathrm{N}}$ from $70.4$ to $73.7$ K within LDA+$U$, whereas it is enhanced from $99.4$ to $122.9$ K within GGA+$U$, respectively.

\begin{table}
\caption{The values of $J_{1}$,$J_{2}$, ..., $J_{6}$  in meV's  obtained from spin spiral calculations and used to fit analytical expression as in \eqref{analy-q}. }
{\def\arraystretch{2}\tabcolsep=1pt
\begin{tabular}{M{1.2cm}M{1.2cm}M{1.2cm}M{1.2cm}M{1.2cm}M{1.2cm}M{1.2cm} }
\hline 
 & $J_{1}$ & $J_{2}$  & $J_{3}$  & $J_{4}$  & $J_{5}$  & $J_{6}$  \\  
\hline 
LDA+$U$ & $-0.285$ & 0.458 & $-0.800$ & $-0.383$ & 0.063 & 0.133 \\ 
\hline 
GGA+$U$ & $\phantom{-}$0.469 & 0.590 & $-0.962$ & $-0.535$ & 0.104 & 0.187\\
\hline 
\end{tabular}
}
\label{j1j6list}\end{table}

\bibliography{apssamp}

\end{document}